\documentclass[12pt,twoside]{article}
\usepackage{xrb2007}
\begin{document}

% select your session by uncommenting the appropriate line
%\session{Jets}
%\session{Jet and Black Hole Binaries}
%\session{Faint Galactic XRB Populations}
%\session{Faint XRBs and Galactic LMXBs}
%\session{Obscured XRBs and INTEGRAL Sources}
%\session{ULXs}
%\session{Extragalactic Populations}
\session{Future Missions and Surveys}
%\session{Population Synthesis}

\shortauthor{Eikenberry}
\shorttitle{F2GCS}

\title{The FLAMINGOS-2 Galactic Center Survey}
\author{Stephen S. Eikenberry}
\affil{Department of Astronomy, University of Florida, 211 Bryant Space
  Science Center, Gainesville, 32611}

\begin{abstract}
Upon commissioning on Gemini South, FLAMINGOS-2 will be one of the most
powerful wide-field near-infrared imagers and multi-object spectrographs ever
built for use on 8-meter-class telescopes.  In order to take best advantage of
the strengths of FLAMINGOS-2 early in its life cycle, the instrument team has
proposed to use 21 nights of Gemini guaranteed time in 3 surveys -- the
FLAMINGOS-2 Early Science Surveys (F2ESS).  The F2ESS will encompass 3
corresponding science themes -- the Galactic Center, galaxy evolution, and
star formation.  In this paper, I review the design performance and status of
FLAMINGOS-2, and describe the planned FLAMINGOS-2 Galactic Center Survey.

\end{abstract}

\section{Introduction}

Multi-object spectroscopy (MOS) is revolutionizing optical astronomy, in
fields as far ranging as abundance studies of globular clusters to the
large-scale structure of the Universe.  Unlike the previously-common
single-object spectrographs, MOS instruments have the capability to observe
tens to hundreds of objects at a single time.  This has enabled large
increases in sample sizes for many studies - often as much as 2 or more orders
of magnitude.

Near-infrared spectroscopy has lagged significantly behind optical
spectroscopy, with the first instruments featuring large-format
(1024x1024-pixel or larger) detector arrays appearing on telescopes in just
the past few years.  In particular, near-infrared MOS have only begun to
appear very recently.  The first fully-cryogenic IR MOS, FLAMINGOS, was
developed at the University of Florida, has seen successful use at the
Gemini South 8-m and MMT 6.5-m telescopes, and is currently in service as a
facility instrument of the Kitt Peak 4-meter telescope (Elston et al. 2002). 

FLAMINGOS-2 (Eikenberry et al. 2006) is a fully cryogenic near-infrared
(0.9-2.5 $\mu m$) wide-field imager and multi-object spectrograph which is
being built by the University of Florida Department of Astronomy for the
Gemini South 8-m telescope on Cerro Pachon, Chile.  FLAMINGOS-2 shares much of
the instrument heritage of FLAMINGOS (Elston et al. 2002), as both a
wide-field imager and MOS.  FLAMINGOS-2 differs from FLAMINGOS primarily in
having optics and opto-mechanical systems optimized for the Gemini telescopes,
providing 0.18-arcsec pixels and a 6.2-arcmin field of view - covering
approximately 6 times the solid angle of FLAMINGOS on the same telescopes.
When commissioned on Gemini 2008, FLAMINGOS-2 will be one of the most
powerful wide-field near-infrared imagers and multi-object spectrographs ever
built for use on 8-meter-class telescopes.

Upon completion of the FLAMINGOS-2 (F2) commissioning work, the F2 instrument
team at the University of Florida (UF) will have access to a number of nights
of guaranteed observing time with F2 on Gemini.  This team consists primarily
of the Principal Investigator, Stephen Eikenberry, and Instrument Team
Scientists, S. Nicholas Raines, Reba Bandyopadhyay, and Anthony Gonzalez.  In
order to take best advantage of the strengths of F2 early in its life cycle,
the instrument team has proposed that the majority of this guaranteed time be
used in 3 surveys - the FLAMINGOS-2 Early Science Surveys (F2ESS). The F2ESS
surveys will encompass 3 corresponding scientific themes - the Galactic
Center, extragalactic astronomy, and star formation.  Each of these surveys
will be carried out by the F2 instrument team in collaboration with groups of
scientists drawn from the Gemini community and elsewhere, with the goal of
maximizing the early scientific return from F2.

In this paper, I will briefly review the designed performance characteristics
of FLAMINGOS-2 as well as the current instrument status.  I will then move on
to a discussion of the FLAMINGOS-2 Galactic Center Survey (F2GCS).

\section{FLAMINGOS-2 Overview \& Status}
\label{sec:F2instrument}

\subsection{Instrument Overview}

FLAMINGOS-2 is an imaging spectrometer for use at the f/16 telescope following
the collimator to produce a reimaged focal surface on the detector array with
2048x2048 18$\mu m$ pixels. A combination of filters and grisms are placed
near the pupil for broad- and narrow-band imaging and moderate-resolution
spectroscopy. A pupil mask reduces excess thermal emission from the telescope.
The imaging mode field will form an inscribed circle on the detector.
FLAMINGOS-2 may also be fed with a slower (f/30) beam provided by the Gemini
Multi-Conjugate Adaptive Optics (MCAO) system.  In spectroscopic mode, a
selection of 9 MOS plates and 3 long slits mask off-target locations in the
focal plane, passing target light through the collimator to a selectable grism
inserted into the beam after the pupil.  The grism disperses the incident
light, which is reimaged as a spectrum on the detector array by the camera
optics.  We present the basic optical performance requirements for FLAMINGOS-2
below.

\
\ 

\begin{tabular}{ll}
%    \toprule
    {\bf Parameter} & {\bf Value} \\
%    \midrule
    Wavelength Range  & $0.9 - 2.5 \mu$m \\
    Imaging field of view  & 6.2-arcmin circular \\
    Pixel scale  & $0.180 \pm 0.002$ arcsec \\
    Detector & HAWAII-2 ($2048 \times 2048$) pix \\
    MOS field of view & $6 \times 2$-arcmin \\
    MOS multiplex gain & up to $\sim 100$ targets \\
    Low-res spectroscopy & $R \sim 1300$ JH or HK bands \\
    High-res spectroscopy & $R \sim 3300$ J, H, or K band \\
    MCAO field of view & $3 \times 1$-arcmin \\
    MCAO pixel scale & 90-mas/pixel \\
%    \bottomrule
\end{tabular}

\ 
\ 

\subsection{FLAMINGOS-2 Status}

FLAMINGOS-2 is currently undergoing final full-system testing at the
University of Florida.  The FLAMINGOS-2 instrument team at the University of
Florida includes Steve Eikenberry (PI), Reba Bandyopadhyay, Greg Bennett,
Richard Corley, Skip Frommeyer, Anthony Gonzalez, Kevin Hanna, Rick Herlevich,
David Hon, Jeff Julian, Roger Julian, Toni Marin, Charlie Murphey, Nick
Raines, William Rambold, David Rashkind, Craig Warner, and the late Richard
Elston.  The FLAMINGOS-2 On-Instrument Wavefront Sensor was developed at the
Herzberg Institute of Astrophysics in Canada by a team including Brian Leckie,
Rusty Gardhouse, Jennifer Dunn, Murray Fletcher, Bob Wooff, and Tim Hardy.

As of this writing, FLAMINGOS-2 has been fully integrated in the laboratory at
the University of Florida, and has successfully demonstrated high-performance
operation in all of its major modes. It will be shipped to Gemini South in
2008 for on-telescope commissioning.  In addition to the instrument hardware
and control software, the UF team has developed a data pipeline tool
for FLAMINGOS-2 called the Florida Analysis Tool Born Of Yearning for
high quality scientific data (FATBOY).  FATBOY combines Python
scripting and code with PyRAF calls to provide imaging, long-slit
spectroscopy, and MOS spectroscopy data reduction capabilities in a rapid,
automated manner.  It has been extensively tested with FLAMINGOS data, and
is currently in scientific use at UF.

\section{The F2 Galactic Center Survey}

The FLAMINGOS-2 Galactic Center Survey (F2GCS) portion of the three F2ESS
surveys is focused on the unusual properties of stars, gas, and black holes at
the center of the Milky Way.  The key goals of the F2GCS are to study and
identify the unuusal population of Chandra-identified X-ray sources at the
Galactic Center, and to use the star formation history of this region to probe
the physics of the ``bulge/black-hole connection'' in galaxies.  This survey
is being led by team co-leads Steve Eikenberry (UF) and Bob Blum (NOAO).
Other team members include F. Baganoff (MIT), F. Bauer (Columbia),
R. Bandyopadhyay (UF), D. Crampton (HIA), C. Dewitt (UF), A. Gonzalez (UF),
M. Muno (Caltech), K. Olsen (NOAO), N. Raines (UF), and K. Sellgren (Ohio
State).

\subsection{X-ray Sources in the F2GCS}

The Galactic Center is a wonderful and mysterious place in our local Universe.
It contains a supermassive black hole in Sgr A*, loads of massive stars and
clusters and, importantly, more than $2000$ identifiable X-ray point sources
in its central region (Muno et al. 2006).  These X-ray sources are unusual in
their properties, being both faint (typical luminosities $L_x < 10^{34} \ {\rm
  ergs \ s^{-1}}$) and spectrally hard.  These properties and their number
density at the Galactic Center are not compatible with other known source
populations in the Galaxy.

Historically, much of the information on X-ray binary source populations in
the Galaxy has come from studies of optically- and infrared-identified
counterparts.  From them, one can determine the donor star type and thus a
rough mass estimate.  Furthermore, optical/IR studies can frequently reveal
variations in brightness and/or wavelength to determine the binary period and
mass function of the system -- two critical parameters for assessing the
nature of the underlying sources.  However, this is a non-trivial task in the
Galactic Center region.  First of all, the high reddening ($A_V \sim 20-40$
mag) makes optical observations highly impractical, so that only infrared
techniques are efficient for these studies.  Secondly, the fields are highly
crowded.  Virtually every Chandra X-ray source has an IR counterpart candidate
within $\sim 1$-arcsec at $K<16$ mag -- but our statistical analyses indicate
that $\sim 85 \%$ of these candidates are spurious chance superpositions.
Thus, we need to sort the ``wheat'' from the ``chaff''.

FLAMINGOS-2 presents the perfect tool for carrying this out.  IR spectroscopy
of potential targets can separate out many false positives, in that X-ray
binaries often carry spectral signatures of accretion -- particularly emission
lines such as Br$\gamma$, HeI, and HeII (e.g. Mikles et al. 2006).  Observing
all 2000 IR candidates should yield $\sim 300$ newly-identified X-ray binaries
in the Galactic Center region -- increasing the number of such optical/IR
identifications in this region by 2 orders of magnitude and effectively
doubling the entire Galactic sample of optical/IR X-ray binary counterparts.
With previous instruments, such observations would take about 1 hour per
target on an 8-meter-class telescope -- or roughly 250 nights of Gemini time
(!!).  However, the massive multiplex gain of FLAMINGOS-2 will allow us to
accomplish this task in $\sim 5-7$ nights of observation -- an eminently
feasible task. 

\subsection{Star Formation History and the F2GCS}

While the F2GCS will find many X-ray source counterparts in a relatively short
amount of time, the fundamental efficiency of searching will only be $\sim 15
\%$, as set by the expected ``false coincidence'' rate (see above).  However,
we will not simply ``waste'' the resulting $\sim 1700$ IR spectra.  Rather,
these HK $R \sim 1300$ spectra will be combined with another $\sim 3000$ stars
selected from our pre-imaging survey to provide nearly $5000$ spectra of stars
in this field -- primarily Red Giant Branch (RGB) stars.  This will produce a
master catalog of many such stars in this field, and the spectra will cover
both the $H/K$ ``steam'' bands and the CO absorption bands.  Combining
measurements of these features will allow us to measure the luminosity class
and extinction for each individual star.  Combining these with photometry
(already obtained from the CTIO 4-meter telescope and ISPI instrument -- see
Dewitt et al., these proceedings) will provide $M_{bol}$ and $T_{eff}$ for
each star.  This in turn places them on a Hertzsprung-Russell diagram.

Blum et al. (2003) used a similar approach based on 75 stars
to constrain the star formation history of the Galactic Center
region. The F2GCS will increase the sample size for this work by nearly
2 orders of magnitude (!).  Just as importantly, the F2GCS will
reach much ($\Delta K \ \sim 5$ mag) fainter than the previous work,
and thus be dominated by stars for which systematic errors due
to atmospheric spectral variations are minimized.  In this way, we
can constrain the star formation history of the Galactic Center over
the past 4 billion years.  Since these stars are effectively
``fossil tracers'' of the motion of star-forming gas through the region
over this long time, we should be able to link this flow to the
mass evolution history of the super-massive black hole in Sgr A*.


\begin{thebibliography}

\bibitem{} Blum, R. D., Ramírez, Solange V., Sellgren, K., Olsen, K. 2003, ApJ, 597, 323
\bibitem{} Eikenberry, Stephen, Elston, Richard, Raines, S. Nicholas, Julian, Jeff, Hanna, Kevin, Hon, David, Julian, Roger, Bandyopadhyay, R., Bennett,  J. Greg, Bessoff, Aaron, Branch, Matt, Corley, Richard, Eriksen, John-David,  Frommeyer, Skip, Gonzalez, Anthony, Herlevich, Michael, Marin-Franch,  Antonio, Marti, Jose, Murphey, Charlie, Rashkin, David, Warner, Craig,  Leckie, Brian, Gardhouse, W. Rusty, Fletcher, Murray, Dunn, Jennifer, Wooff,  Robert, Hardy, Tim 2006, Proc. of SPIE, 6269, 39
\bibitem{} Elston, R. 2002, Proc. of SPIE, 4841, 1611
\bibitem{} Mikles, Valerie J., Eikenberry, Stephen S., Muno, Michael P., Bandyopadhyay, Reba M., Patel, Shannon 2006, ApJ, 651, 408
\bibitem{} Muno, M. P., Bauer, F. E., Bandyopadhyay, R. M., Wang, Q. D. 2006, ApJS, 165, 173

\end{thebibliography}
\end{document}